\documentclass[epj]{svjour}
\usepackage{graphics}
\usepackage{cite}
\usepackage{multirow}
\usepackage{booktabs}
\usepackage{latexsym} 	
\usepackage{amsmath}  	
\usepackage{amsbsy}
\usepackage{epsfig}     
\usepackage{graphicx}
\usepackage{latexsym}
\usepackage{euscript}
\usepackage{amsfonts}
\usepackage{mathtools}
\usepackage{simplewick}
\usepackage{amsmath}
\usepackage{amssymb}
\usepackage{dsfont}
\usepackage{slashed}
\usepackage{hyperref}

\newcommand{\be}{\begin{equation}}
\newcommand{\ee}{\end{equation}}
\newcommand{\bea}{\begin{eqnarray}}
\newcommand{\eea}{\end{eqnarray}}
\newcommand{\bit}{\begin{itemize}}
\newcommand{\eit}{\end{itemize}}
\newcommand{\bmp}[1]{\begin{minipage}{#1cm}}
\newcommand{\emp}{\end{minipage}}

\newcommand{\bra}{\langle}
\newcommand{\ket}{\rangle}

\newcommand{\im}{{\mathrm{Im}}}

\newcommand{\pv}{{\mathbf p}}
\newcommand{\kv}{{\mathbf k}}

\newcommand{\xv}{{\mathbf x}}
\newcommand{\half}{\frac{1}{2}}

\newcommand{\eps}{\epsilon}

\newcommand{\om}{\omega}

\newcommand{\bean}{\begin{eqnarray*}}
\newcommand{\eean}{\end{eqnarray*}}
\newcommand{\nn}{\nonumber}
\newcommand{\id}{1\!\!1}
\newcommand{\vecB}{{\mathbf B}}

\newcommand{\vecp}{{\mathbf p}}

\newcommand{\vecnul}{{\mathbf 0}}
\begin{document}

\title{Electrical conductivity of the quark-gluon plasma: perspective from lattice QCD}

\author{Gert Aarts\thanks {Email: g.aarts@swan.ac.uk} 
\and Aleksandr Nikolaev\thanks {Email: aleksandr.nikolaev@swan.ac.uk}
} 

\institute{Department of Physics, College of Science, Swansea University, Swansea SA2 8PP, United Kingdom}

\date{%Received: date / Revised version: date
August 27, 2020
}

\abstract{
A discussion on the electrical conductivity of the quark-gluon plasma as determined by lattice QCD is given. After a reminder of basic definitions and expectations, various methods for spectral reconstruction are reviewed, including the use of Ans\"atze and sum rules, the Maximum Entropy and Backus-Gilbert methods, and Tikhonov regularisation. A comprehensive overview of lattice QCD results obtained so far is given, including a comparison of the different lattice formulations. 
A noticeable consistency for the conductivities obtained is seen, in spite of the differences in the lattice setups and spectral reconstruction methods.
It is found that in the case of quenched QCD little temperature dependence of $\sigma/T$ is seen in the temperature range investigated, while for QCD with dynamical quarks a reduction of $\sigma/T$ in the vicinity of the thermal crossover is observed, compared to its value in the QGP. Several open questions are posed at the end.  
\\
\vspace*{-0.3cm}\\
{\scriptsize Invited contribution to the EPJA topical issue ``Theory of hot matter and relativistic heavy-ion collisions (THOR)".}
\vspace*{-0.5cm}\\
 % end of PACS codes
} %end of abstract

\maketitle

%-----------------------------------------------------------------------------------------------------

\section{Introduction}
\label{sec:intro}

The experimental heavy-ion programmes at the Relativistic Heavy Ion Collider (RHIC) and the Large Hadron Collider (LHC), and, in the near future, at the Nuclotron-based Ion Collider fAcility (NICA) and the Facility for Antiproton and Ion Research (FAIR), offer exciting probes into the dynamics of strongly interacting matter under extreme conditions. The relation with the underlying theory, Quantum Chromodynamics, is established via phenomenology, which permits a connection between quantities computable from first principles, such as the equation of state, and measurable observables in the experiments.

In this contribution we focus on one such quantity, the electrical conductivity $\sigma$ of the quark-gluon plasma (QGP). As we will review in the next section, the usual definition of the conductivity, employing the Kubo formula, relates it to a specific limit of Green's functions in quantum field theory, allowing for a computational formulation using, e.g., lattice QCD {\em in principle}. On the other hand, the conductivity plays a role in charge transport, particle production and the time evolution of electromagnetic fields generated in heavy-ion collisions, see e.g.\ Refs.\ 
\cite{Deng:2012pc,Hirono:2012rt,Tuchin:2013ie,Yin:2013kya,Kharzeev:2015znc,Pratt:2019pnd}
and references therein, emphasising its phenomenological relevance.

On the theoretical side, the conductivity can be computed using a variety of methods, ranging from Feynman diagrams at weak coupling \cite{Aarts:2002tn,Gagnon:2006hi}
 and kinetic theory in QCD \cite{Arnold:2000dr,Arnold:2003zc} or effective models \cite{Cassing:2013iz,Marty:2013ita} to holographic methods at strong coupling \cite{Policastro:2002se,Teaney:2006nc}. Here we will focus on the results obtained using numerical simulations of QCD discretised on the lattice, as a first-principle tool to access nonperturbative information in  the vicinity of the deconfinement transition.

So far there are ${\cal O}(10)$ papers which have attempted to compute the conductivity on the lattice \cite{Gupta:2003zh,Aarts:2007wj,Ding:2010ga,Francis:2011bt,Brandt:2012jc,Amato:2013naa,Aarts:2014nba,Brandt:2015aqk,Ding:2016hua,Astrakhantsev:2019zkr}. 
These papers differ substantially in detail, partly indicating the increase in available computing power over the past 15 years or so. For instance, there are simulations with $N_f=0$ flavours (quenched QCD), and $N_f=2$ and $2+1$ dynamical flavours; with quarks heavier than in nature or at the physical point; using a continuum extrapolation or at fixed lattice spacing; with isotropic and anisotropic ($a_\tau\ll a_s$) lattices, etc (a detailed comparison is given in Sec.\ \ref{sec:lattice}). Importantly, the methods used to extract the conductivity from the Euclidean lattice correlators differ substantially as well, and include the use of Ans\"atze, Bayesian approaches such as the Maximum Entropy Method, and other regularisations.
Despite these differences, a consistent picture is seen to emerge, with approximate agreement between simulations with either dynamical quarks or in the quenched case. The aim of this review is to give a comprehensive overview of what has been obtained so far and provide a comparison of the results. For completeness, we note here that we restrict ourselves to the conductivity in the case of light quarks; we will not discuss heavy-quark diffusion~\cite{Meyer:2010tt,Banerjee:2011ra,Francis:2015daa} and neither other transport coefficients such as the shear~\cite{Meyer:2007ic,Astrakhantsev:2017nrs,Pasztor:2018yae} and bulk viscosities~\cite{Meyer:2007dy,Astrakhantsev:2018oue}.

This paper is organised as follows. In the following section we present some basic expressions relating the conductivity to various Green's functions, notably the spectral function and the corresponding Euclidean correlator, using the Kubo relation. Some general remarks on expectations at high temperature and the so-called transport peak are given as well. In Section \ref{sec:recon} we discuss the various approaches that have been employed to reconstruct the spectral function and extract the conductivity, given a numerically determined Euclidean correlator. An overview of available lattice results is given in Section \ref{sec:lattice}, including a comparison between the values of $\sigma$ obtained so far. Some related developments are summarised in Section \ref{sec:related}. The final section contains a summary, including some open questions. In the Appendix some well-known relations between the various Green's functions are collected. 

%-----------------------------------------------------------------------------------------------------

\section{Kubo formula and spectral function}
\label{sec:kubo}

The electromagnetic current in QCD receives contributions from all quark flavours and reads
\be
j_\mu^{\rm em}(x) = \sum_{f=1}^{N_f} (eq_f) j_\mu^f(x), \quad
j_\mu^f(x) = \bar\psi^f(x)\gamma_\mu\psi^f(x).
\ee
Here $q_f$ denotes the fractional charge of the quark ($2/3$ or $-1/3$) and $e$ the elementary charge. We restrict the discussion to light quarks, with $N_f=2$ or $2+1$. The current is hermitian, ${j_\mu^{\rm em}}^\dagger(x)=j_\mu^{\rm em }(x)$.

The electrical conductivity $\sigma$ indicates the linear relationship between the current density and an electric field, $j_i^{\rm em}=\sigma E_i$, according to Ohm's law. Using linear-response theory, it can be related to the current-current correlator in thermal equilibrium, in absence of the external electric field, see e.g.~Ref.~\cite{Arnold:2000dr}. More precisely, the conductivity is proportional to the slope of the current-current spectral function in thermal equilibrium,
\be
\rho_{\mu\nu}^{\rm em}(\om,\pv) = \int d^4x\, e^{i\om t-\pv\cdot\xv} \bra [j_\mu^{\rm em}(t,\xv), j_\nu^{\rm em}(0,\vecnul)]\ket,
\ee
at vanishing energy and momentum, i.e.,
\be
\label{eq:cond}
\sigma = \frac{1}{6}\frac{\partial}{\partial\om}\rho^{\rm em}_{ii}(\om,\vecnul)\Big|_{\om=0}.
\ee
Here the summation over spatial components, $i=1,2,3$, is understood. The current-current spectral function is the expectation value of the commutator of the electromagnetic current, evaluated at temperature $T$.

The conductivity is closely related \cite{Arnold:2000dr} to the charge diffusion coefficient $D$, according to the Einstein relation,
\be
\sigma = \chi_Q D,
\ee
where $\chi_Q$ is the charge susceptibility,
\be
\chi_Q = \frac{1}{TV} \left\bra (Q-\bra Q\ket)^2\right\ket.
\ee
Here $V$ is the spatial volume and $Q$ is the total charge, i.e.\ the volume integral of $j_0^{\rm em}(x)$. 

Some well-known relations between the spectral function and other Green's functions are given in Appendix \ref{app:A}. In particular, the spectral function is related to the Euclidean correlator,
\be
\label{eq:Gem}
G_{\mu\nu}^{\rm em}(\tau, \xv) = \bra j_\mu^{\rm em}(\tau, \xv) j_\nu^{\rm em}(0,\vecnul)\ket,
\ee
via the standard relation [see Eq.\ (\ref{eq:AG})]
\be
\label{eq:Grho}
G_{\mu\nu}^{\rm em}(\tau,\pv) =  \int_0^\infty \frac{d\om}{2\pi}\, K(\tau,\om)\rho_{\mu\nu}^{\rm em}(\om, \pv),
\ee
with the kernel
\be
\label{eq:kernel}
K(\tau,\om) =  \frac{\cosh[\om(\tau-1/2T)]}{\sinh(\om/2T)}.
\ee
Here the Euclidean time $0\leq \tau<1/T$.
The question of computing the conductivity on the lattice therefore boils down to numerically computing Eq.~(\ref{eq:Gem}), inverting Eq.~(\ref{eq:Grho}) and extracting the slope according to Eq.~(\ref{eq:cond}).
From now we work at vanishing spatial momentum and drop the $\pv$ dependence. When prefactors involving $eq_f$ are dropped, the superscript `em' is omitted as well.

\begin{figure}[t]
\begin{center}
    \vspace*{-0.4cm}
    \includegraphics[width=0.2\textwidth]{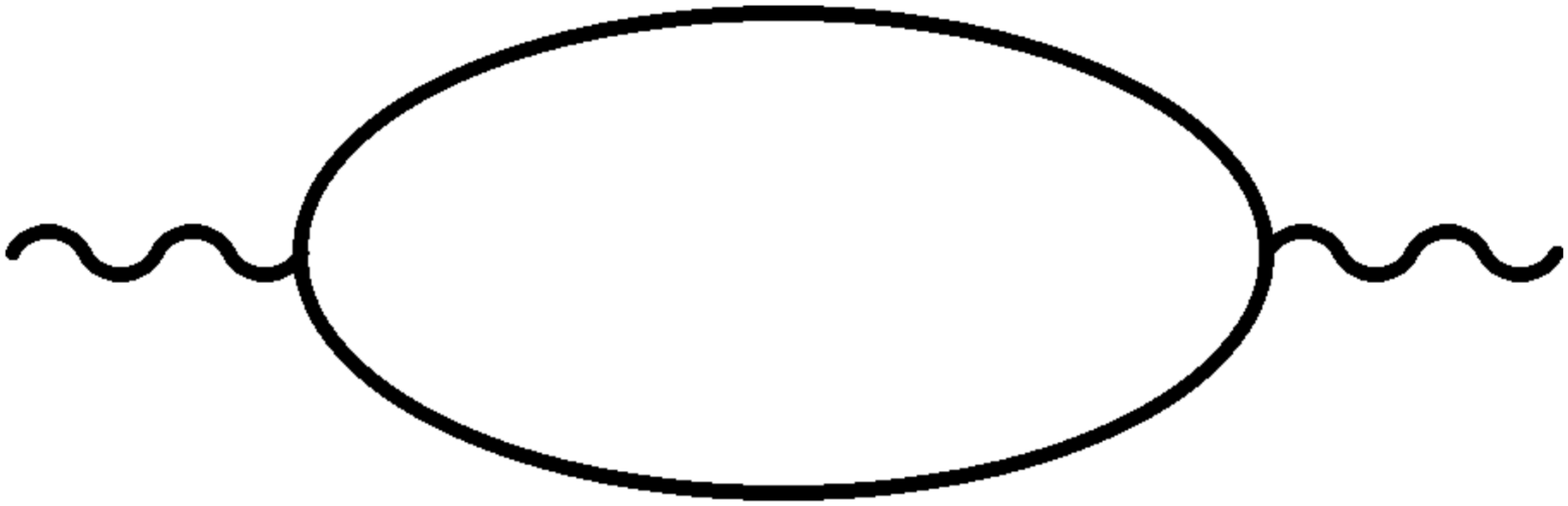}
    \vspace*{-0.4cm}
    \caption{One-loop contribution to the current-current spectral function.}
    \label{fig:oneloop}
\end{center}
\end{figure}

In order to prepare for the discussion of the lattice QCD results below, it is useful to recall what can be expected at very high temperature, where QCD is weakly coupled. At leading (zeroth) order in perturbation theory (see Fig.~\ref{fig:oneloop}), the spectral function, for a single flavour with mass $m$, reads \cite{Aarts:2005hg}\footnote{Note there is a typo in the last line of Eq.\ (19) in Ref.\ \cite{Aarts:2005hg}: $a^{(1)}_H + a^{(2)}_H$ should read $a^{(1)}_H + a^{(3)}_H$.}
\bea
\rho_{ii}(\om) &=& 2\pi N_c I\om\delta(\om) \nn\\
&&+ \frac{N_c}{2\pi} \theta(\om^2-4m^2)\sqrt{\frac{\om^2-4m^2}{\om^2}} \nn\\
&&\times \left(\om^2+2m^2\right)\left[1-2n_F\left(\frac{\om}{2}\right)\right],
\label{eq:rho-free}
\eea
with $N_c=3$. Here $n_F(\om)=1/[\exp(\om/T)+1]$ is the Fermi-Dirac distribution.
The quantity $I$ in the first term reads
\be
I = -4 \int \frac{d^3k}{(2\pi)^3}\, n_F'(\om_\kv) \frac{k^2}{\om_\kv^2},
\ee
with $\om_\kv=\sqrt{\kv^2+m^2}$. For massless quarks this evaluates as
\be
I\big|_{m=0} = \frac{T^2}{3}.
\ee
Since
\be
1-2n_F\left(\frac{\om}{2}\right) = \tanh\left(\frac{\om}{4T}\right),
\ee
the spectral function is odd, $\rho_{ii}(-\om)=-\rho_{ii}(\om)$, as it should be. Below we take $\om\geq 0$.
The corresponding Euclidean correlator reads, in the massless limit \cite{Aarts:2005hg},
\be
G_{ii}(\tau)= N_c T^3 \left[ \frac{1}{3}+ \frac{3u+u\cos(2u)-2\sin(2u)}{\sin^3(u)} \right],
\ee
where $u=2\pi T(\tau - 1/2T)$. The first (constant) term comes from the first term in Eq.~(\ref{eq:rho-free}); the $\tau$ dependent term from the second one. At the midpoint, $\tau=1/2T$, the contributions from both terms are comparable, $G_{ii}(\tau=1/2T) = N_cT^3(1/3+2/3)=N_cT^3$.

Let us now discuss the two contributions in Eq.~(\ref{eq:rho-free}) in more detail. We start with the second one. This term arises from the cut of the one-loop polarisation diagram in Fig.~\ref{fig:oneloop} corresponding to a decay process, with $\om=\om_\kv+\om_{\pv+\kv}$ (with $\kv$ the loop momentum and $\pv=\vecnul$), which is permissible once $|\om|>|2m|$. It contains the vacuum contribution, increasing as $\om^2$ at large $\om$, and a thermal contribution, leading to Pauli blocking, which is exponentially suppressed at large energies. Note that this term does not contribute to the conductivity; in the massless limit it increases as $\om^3$ as $\om\to 0$. We will refer to this term as the continuum or perturbative contribution.

The first term, with $\om\delta(\om)$, is only present at nonzero temperature and arises from the cut corresponding to scattering with particles in the heatbath, $\om_\kv+\om =\om_{\pv+\kv}$, in the limit that $\pv\to \vecnul$. Extracting the conductivity from this term yields infinity, reflecting the fact that for free particles the mean free path and hence the conductivity diverges. Interactions make the mean free path finite, due to scattering in the plasma. The result is that the $\delta$ function in Eq.\ (\ref{eq:rho-free}) is smeared out and takes the form of a so-called transport peak,
\be
\label{eq:rhotrans}
2\pi N_c I \om\delta(\om) \to \rho_{\rm trans}(\om) = A_{\rm trans} \frac{\gamma \om}{\om^2+\gamma^2},
\ee
where $\gamma$ is proportional to the collisional scattering width or the inverse mean free path, $A_{\rm trans}(=2N_c I)$ is the overall coefficient, and $\sigma\propto A_{\rm trans}/\gamma$. The origin of the transport peak can be seen in various ways, using e.g.\ Feynman diagrams and pinching poles \cite{Aarts:2002cc,Aarts:2002vx} or kinetic theory \cite{Petreczky:2005nh,Meyer:2011gj}. It should be noted that this form of transport peak is the simplest form encountered. An important observation is that in the case of a narrow transport peak ($\gamma \ll T$), as is the case for weakly-coupled theories, the Euclidean correlator is not sensitive to details of the transport peak but only to its area \cite{Aarts:2002cc}. In the limit that $\om < \Lambda \ll T$, the kernel (\ref{eq:kernel}) simplifies to $2T/\om$, and integrating the transport peak (\ref{eq:rhotrans}) in Eq.\ (\ref{eq:Grho}) yields
\bea
\nn
&& G_{\rm trans}(\tau) = \int_0^{\Lambda} \frac{d\om}{2\pi}\, K(\tau,\om)\rho_{\rm trans}(\om) \\
\label{eq:Gtrans}
&&\sim A_{\rm trans} \int_0^{\Lambda} \frac{d\om}{2\pi}\, \frac{2T}{\om} \frac{\gamma \om}{\om^2+\gamma^2}
\sim \half A_{\rm trans} T,
\eea
where in the last expression the cutoff $\Lambda$ on $\om$ has been removed. The crucial observation is that this expression is independent of $\gamma$ and the Euclidean time $\tau$, indicating the insensitivity of the correlator to narrow transport peaks. In fact, taking $A_{\rm trans}=2N_c I$, its value is the same as in the non-interacting theory. 

So far we have only considered the diagram given in Fig.\ \ref{fig:oneloop}, dressed with gluons and closed loops of sea quarks in the presence of interactions (in lattice QCD this is referred to as the connected contribution). However, when interactions are included, there are also contributions from diagrams with a different topology, namely with two closed fermion loops, connected via gluons. These arise from Wick contractions of the current operator, $j_i = \bar\psi\gamma_i\psi$, with itself. 
In lattice QCD, this is commonly referred to as the disconnected contribution. Perturbatively, they start contributing at ${\cal O}(\alpha_s^3)$ only, and hence are suppressed at very high $T$. 
Another distinction between the connected and disconnected contributions concerns the appearance of the electromagnetic charges. Take for simplicity $N_f=2$ or $3$ degenerate flavours.   
In the connected contribution one finds the sum over charges squared, which is usually denoted as 
\be
\label{eq:Cem}
C_{\rm em} = \sum_f (e q_f)^2  =
\begin{cases}
\frac{5}{9} e^2 \qquad (u, d), \\
\frac{6}{9} e^2 \qquad (u, d, s).
\end{cases}
\ee
In the disconnected contribution on the other hand, we find the square of the sums, 
\be
C_{\rm em}^{\rm disc} = \big(\sum_f e q_f \big)^2  = 
\begin{cases}
\frac{1}{9} e^2 \qquad (u, d), \\
0 \qquad\quad  (u, d, s).
\end{cases}
\ee
Hence it is noted that the disconnected contribution vanishes for three degenerate flavours.

Up to now we have focussed on the expectation at high temperature. At low temperature, in the hadronic phase, the current-current correlator couples to vector mesons, with details depending on the flavour content. A comprehensive discussion in the $N_f=2$ case can be found in Ref.\ \cite{Brandt:2012jc}. In the spectral function one therefore expects bound-state peaks, representing mesonic ground and excited states. As the system is heated, these bound states are expected to dissolve and the high-temperature behaviour to emerge. 

To conclude this section, we note that lattice QCD simulations include quarks scattering with gluons, i.e.\ the electromagnetic field and other charge carriers (leptons) are not included. Hence the conductivity reviewed here indicates the contribution from the strong interaction only, yielding insight into the strongly coupled nature of the quark-gluon plasma.

%-----------------------------------------------------------------------------------------------------

\section{Spectral reconstruction}
\label{sec:recon}

The main problem in extracting the conductivity from numerically determined Euclidean lattice correlators is the inversion problem, see Eq.~(\ref{eq:Grho}). As a reminder, on the lattice temperature is encoded in the compact direction in imaginary time, with circumference $1/T= a_\tau N_\tau$. Here $a_\tau$ is the temporal lattice spacing and $N_\tau$ the number of points in the time direction; Euclidean time is discretised as $\tau/a_\tau=0, \ldots, N_\tau-1$.
Since the correlator $G(\tau)$ is known at a finite number of temporal points and the spectral function $\rho(\om)$ is in principle a continuous function of $\om$, this inversion problem is far from straightforward. To be more precise, due to reflection symmetry, $K(\tau,\om)=K(1/T-\tau,\om)$,  the number of points available for the analysis is on the order of $N_\tau/2$; even after placing an upper limit on the $\om$ interval, such that $0<\om<\om_{\rm max}$, and discretising the finite interval, typically on the order of $N_\om=1000$ points are used to present $\rho(\om)$. Since $N_\om\gg N_\tau$, the inversion problem is ill-posed.
In addition, the focus on the  $\om\to 0$ limit makes the inversion more challenging than for spectral functions in general, when the interest is in frequencies on the order of the temperature or above, as the discussion around Eq.~(\ref{eq:Gtrans}) indicates.

Several methods have been developed to tackle this problem. Here we briefly review the ones applied to the conductivity. It is fair to state that no single method is yet fully robust on its own. Hence it is of interest to compare and contrast the results obtained so far, and seek for (in)consistencies. This will be done in Sec.~\ref{sec:lattice}.

\subsection{Reconstructed correlators}

Before investigating the temperature dependence of the spectral function, we note that the Euclidean correlator (\ref{eq:Grho}) depends on temperature in two ways:
\begin{itemize}
\item via the temperature dependence of the kernel, $K(\tau,\om)$, due to the compact time direction, $0\leq \tau<1/T$;
\item via the temperature dependence of the spectral function, due to changes in the quark-gluon plasma. 
\end{itemize}
The first effect leads to temperature dependence of the correlator even when the spectral function is unchanged. It is important to disentangle this from the sought second (physical) effect due to actual changes in the plasma. This can be investigated using so-called reconstructed correlators \cite{Datta:2003ww,Aarts:2007pk}. Let us suppose a spectral function $\rho(\om;T_0)$ is determined (with some confidence) at a reference temperature $T_0$. 
Assuming that the spectral function is unchanged, a correlator at a different temperature $T$ can then be defined as  
\be\label{eq:G_reconstructed_def}
G_{\rm recon}(\tau, T;T_0) = \int_0^\infty \frac{d\om}{2\pi}\, K(\tau,\om;T) \rho(\om;T_0).
\ee
This construction takes into account the trivial temperature dependence due to the kernel, the first effect above. Comparing this reconstructed correlator with the actual correlator at temperature $T$ allows one to draw conclusions on the second effect, i.e.\ changes in the spectral function due to a change in the physical situation.
A difference between the actual and the reconstructed correlator implies a change in the spectral function (the inverse is not necessarily true).

\subsection{Sum rules}

Exact sum rules are important \cite{Brandt:2012jc} to constrain the current-current spectral function at nonzero temperature. Defining the difference between the finite and zero-temperature spectral function as
\be
\label{eq:Drho}
\Delta \rho(\om,T) \equiv \rho_{ii}(\om,T) - \rho_{ii}(\om,0),
\ee
one finds the sum rule \cite{Brandt:2012jc}, in the thermodynamic limit, 
\be
\label{eq:sumrule}
\int_{-\infty}^\infty \frac{d\om}{\om} \Delta\rho(\om,T)=0.
\ee
Note that the zero-temperature $\om^2$ contribution cancels in the subtraction (\ref{eq:Drho}). Since the Operator Product Expansion (OPE) predicts \cite{CaronHuot:2009ns} that the thermal contribution decays as $(T/\om)^2$ at large $\om$, the integral in the sum rule converges. One may verify that this sum rule indeed holds for free fermions, using Eq.~(\ref{eq:rho-free}).

The sum rule indicates that enhancement of spectral weight at small energies, i.e.\ due to a larger transport peak, should be compensated by a loss of spectral weight elsewhere. 
Since the sum rule is exact, it should be satisfied by reconstructed spectral functions on the lattice, where it can be implemented as a check or a constraint.
This sum rule, and two additional ones, are further analysed in Refs.~\cite{Gubler:2016hnf, Gubler:2017qbs}.

\subsection{Ans\"atze}

The first step to resolve the ill-posedness of the inversion is to reduce the number of parameters needed to model the spectral function. The easiest way to do so is by providing an Ansatz for $\rho(\om)$ with less fit parameters than data points. The downside is that this introduces an obvious bias, which is difficult to avoid. Moreover, since the spectral function is expected to behave in quite a different manner in the low- and high-temperature phases, the Ansatz has to be sufficiently rich to capture this. Some features to be included are
\begin{itemize}
    \item a transport peak at small $\om$, with in particular a linear slope in $\om$; 
    \item continuum ($\om^2$) contribution at high $\om$, possibly modified by lattice artefacts \cite{Karsch:2003wy,Aarts:2005hg};
    \item at least one bound-state peak in the low-temperature phase, to represent the vector meson.
\end{itemize}
Refs.~\cite{Ding:2010ga,Francis:2011bt,Ding:2016hua} employ an Ansatz combining a transport peak and the expected perturbative continuum behaviour (for massless quarks) in the deconfined phase,   
\be
\rho(\om) = \rho_{\rm trans}(\om) + \rho_{\rm pert}(\om),
\ee
where $\rho_{\rm trans}(\om)$ is given in Eq.\ (\ref{eq:rhotrans}) and 
\be
\label{eq:rhopert}
\rho_{\rm pert}(\om) = \frac{3}{2\pi} A_{\rm pert} \om^2 \left[ 1-2n_F\left(\frac{\om}{2}\right)\right],
\ee
c.f.\ Eq.~(\ref{eq:rho-free}).
The three temperature-dependent parameters are the coefficients $A_{\rm trans, pert}$ and the width $\gamma$ of the transport peak. Note that $A_{\rm pert}=1$ for free fermions; it parametrizes deviations from a free spectral function at large energies. As stated, the functional form of this Ansatz is the combination of two functions. Modifying the transport peak to a flat featureless function, as seen e.g.\ in holography \cite{Teaney:2006nc}, Ref.\ \cite{Ding:2016hua} finds that the data may not have the resolution to differentiate between these two shapes. This is incorporated in the systematic uncertainty of the final quoted result for $\sigma$ \cite{Ding:2016hua}.

Ref.\ \cite{Brandt:2012jc} employs a related Ansatz for the subtracted spectral function, $\Delta \rho(\om, T)$, defined in Eq.\ (\ref{eq:Drho}). In addition to the transport peak and the continuum contribution, this Ansatz also includes a bound-state peak. It reads
\be
\Delta\rho(\om) = \rho_{\rm trans}(\om) + \Delta\rho_{\rm pert}(\om) + \rho_{\rm bound}(\om),
\ee
with the new term 
\be
\rho_{\rm bound}(\om) = A_{\rm bound}\frac{2 g_B \tanh(\om/T)^3}{4(\om-m_B)^2 + g_B^2}.
\ee
Here $m_B, g_B$ and $A_{\rm bound}$ indicate the mass,  width and strength of the bound state. The factor $\tanh(\om/T)^3$ ensures the contribution does not contribute to the conductivity in the $\om\to 0$ limit and decays as $1/\om^2$ at large $\om$, as predicted by the OPE \cite{CaronHuot:2009ns}.
It is also noted in Refs.\ \cite{Burnier:2012ts,Brandt:2012jc} that the transport peak (\ref{eq:rhotrans}) in fact violates this condition. Hence it is proposed \cite{Brandt:2012jc} to modify it as
\be
\rho_{\rm trans, mod}(\om) = A_{\rm trans} \frac{T\tanh(\om/T) \gamma}{\om^2+\gamma^2}\,, 
\ee
which still has linear behaviour at small $\om$ but decays as $1/\om^2$ at large $\om$. 
Finally, the main reason for the subtraction is to eliminate the zero-temperature $\om^2$ contribution, 
\be
\Delta\rho_{\rm pert}(\om) = \rho_{\rm pert}(\om;T) - \rho_{\rm pert}(\om;T=0),
\ee
which eliminates the ``1'' in Eq.\ (\ref{eq:rhopert}). This allows the analysis to focus on frequencies on the order of the temperature, without being overwhelmed by the $\om^2$ term.
Overall, the number of parameters ($A_{\rm trans, pert, bound}, m_B, g_B, \gamma$) to be fitted is quite large, which is carried out by fixing them in steps, while satisfying the sum rule (\ref{eq:sumrule}).
A reduced model, using only $\rho_{\rm trans} + \rho_{\rm pert}$, is employed as well. A variation of this Ansatz is also used in Ref.\ \cite{Brandt:2015aqk}, replacing the bound state by a delta-function and introducing explicit thresholds for the various terms.

\subsection{Maximum Entropy Method}

The Ans\"atze described above have to incorporate a wide range of physics input (bound states, transport peak, continuum contribution), each of which is defined by a number of parameters, making the fit highly nonlinear and depending on the choice of model functions. It is therefore desirable to use model-independent reconstruction methods for the spectral function.   
It will be necessary to regularise standard minimisation procedures, due to the ill-posedness of the inversion. Before proceeding, we note the possibility to rescale the kernel and the spectral function, 
\be
\label{eq:rescale}
K(\tau, \omega) \to f(\omega) K(\tau, \omega), \qquad \rho(\omega) \to \rho(\omega)/f(\omega),
\ee
leaving the product unchanged, to stabilise the inversion. A common rescaling is to use $f(\om)=\om$, to resolve the $1/\om$ divergence in the kernel as $\om\to 0$ \cite{Aarts:2007wj}.

There is still a requirement to reduce the number of parameters to be determined, to make the inversion solvable. There is considerable freedom to do so. 
For instance, in the Maximum Entropy Method (MEM), and in particular Bryan's method \cite{Bryan}, the (rescaled) spectral function is parametrised as \cite{Aarts:2007wj,Asakawa:2000tr}
\be
\label{eq:rho-default}
\frac{\rho(\om)}{\om} = \frac{m(\om)}{\om} \exp \sum_{i=1}^{N_{\rm coeff}} c_i u_i(\om),
\ee
with $u_i(\om)$ $(i=1,\ldots, N_{\rm coeff})$ an orthogonal but incomplete set of basis functions,
\be
\int_0^\infty \frac{d\om}{2\pi} \, u_i(\om)u_j(\om) = \delta_{ij}.
\ee
The reduction follows since $N_{\rm coeff} \sim N_\tau/2 \ll N_\om$. The form (\ref{eq:rho-default}) is motivated by positivity, $\rho(\om)/\om\geq 0$.  In MEM, $m(\om)$ is referred to as the default model, see below. The conductivity is now determined by
\be
\sigma \sim m^\prime(0) \exp\sum_i c_i u_i(0),
\ee
and hence depends on all coefficients and the default model.

In MEM the coefficients $c_i$ are determined by constructing the most probable spectral function as defined by the extremum of the conditional probability $P(\rho | DH)$ \cite{Asakawa:2000tr}. Here $D$ indicates the data and $H$ additional prior knowledge. The method relies on Bayes' theorem, 
\be
\label{eq:MEM_def}
P(\rho|DH) = \frac{P(D|\rho H) P(\rho|H)}{P(D|H)},
\ee
where $P(A|B)$ stands for the conditional probability of $A$ given $B$. In this expression, $P(D|\rho H)$ is the likelihood function, $P(\rho|H)$ the prior probability, and $P(D|H)$ a normalisation. While the likelihood function, $P(D|\rho H) = e^{-L(\rho)}$, is familiar from standard $\chi^2$-minimisation, the prior probability contains an entropy-like term, 
\be
\label{eq:MEM_PrhoI_def}
P(\rho|H) = e^{\alpha S(\rho)},
\ee
with
\be
S(\rho) = \int_0^{\infty} \frac{d\omega}{2\pi}\,\left[ \rho(\omega) - m(\omega) - \rho(\omega) \ln\frac{\rho(\omega)}{m(\omega)} \right],
\ee
giving the method its name.
The conditional probability now reads
\be
\label{eq:MEM_final}
P(\rho|DH) \propto e^{- L(\rho) + \alpha S(\rho)},
\ee
with $\alpha$ determining the balance between the two terms. While at $\alpha=0$ the method reduces to a standard fitting procedure, in absence of any data the probability is extremised when $\rho(\om)=m(\om)$, yielding the default result. 
For further details on MEM we refer to Ref.\ \cite{Asakawa:2000tr}.

MEM has been applied to the conductivity in Refs.~\cite{Gupta:2003zh,Aarts:2007wj,Ding:2010ga,Francis:2011bt,Amato:2013naa,Aarts:2014nba}. We note here that the basis functions $u_i(\om)$ are obtained via a singular value decomposition (SVD) of the kernel $K(\tau,\om)$, when viewed as a $N_{\rm coeff}\times N_\om$ matrix, where $N_{\rm coeff}\lesssim N_\tau/2$, linking the size of the set of basis functions to the temporal extent.  This is a limitation for small $N_\tau$ and has motivated the use of anisotropic lattices to increase the number of data points and basis functions at a given temperature, see Ref. \cite{Amato:2013naa} and especially Ref.\ \cite{Aarts:2014nba} for a systematic study.
While at first sight the choice of default model appears to play an important role in the formulation, in practice it has been found that the dependence on $m(\om)$ is quite mild \cite{Aarts:2014nba}. 
The main systematic uncertainty enters via the formulation of the method, such as the choice of the prior probability, which can be addressed e.g.\ by a comparison with other independent approaches. 

\subsection{Backus-Gilbert method}
\label{sec:BG_method}

The Backus-Gilbert method is such an independent approach, designed for solving linear ill-defined problems with controllable regularisation and systematic uncertainty. Rather than reconstructing the entire spectral function $\rho(\om)$, it aims to represent it by an estimator
\begin{eqnarray}
\label{eq:rho_in_terms_of_resolution}
\widehat\rho(\omega_0) = \int_0^\infty d\om\, \delta(\omega_0, \omega) \rho(\omega),
\end{eqnarray}
where $\delta(\omega_0, \omega)$ is called the resolution function, which should be narrowly peaked around $\om_0$ and normalised,
\be
\label{eq:delta_resolution_norm}
\int_0^\infty d\omega\,\delta(\omega_0, \omega) = 1,
\ee
similar to a delta function. 
Ideally one wants to make the resolution function as narrow as possible, for given correlator and kernel.
For this purpose the following linear Ansatz is assumed \cite{Brandt:2015aqk,Astrakhantsev:2019zkr}
\begin{equation} \label{eq:delta_resolution_setup}
\delta(\omega_0, \omega) = \sum_{n=1}^{N_\tau - 1} q_n(\omega_0) K(\tau_n, \omega).
\end{equation}
The functions $q_n(\omega_0)$ are found by minimising the second moment of the resolution function squared,
\be
\label{eq:delta_resolution_2nd_moment}
\Gamma_{\om_0} = \int_0^\infty d\omega\, (\omega - \omega_0)^2 \delta^2(\omega_0, \omega),
\ee
which should effectively minimise its width. Combining the equations above, one obtains the solution (summation over repeated indices implied)
\be \label{eq:q_solution}
q_n(\omega_0) = \frac{W(\omega_0)^{-1}_{nm} R_m}{R_k W(\omega_0)^{-1}_{kl} R_l},
\ee
in terms of
\bea \label{eq:W}
W(\omega_0)_{nm} &=& \int_0^\infty d \omega\, (\omega - \omega_0)^2 K(\tau_n, \omega) K(\tau_m, \omega),  \\
R_n &=& \int_0^\infty d \omega\, K(\tau_n, \omega)\,.
\eea
Note that since the kernel $K(\tau,\om)$ is symmetric around the midpoint, the number of independent basis functions $q_n$ is limited by $N_\tau/2$. In general, the larger the number of available time slices, the narrower resolution functions can be constructed.

The spectral function can now be estimated by combining Eqs.\ (\ref{eq:rho_in_terms_of_resolution}, \ref{eq:delta_resolution_setup}) and the definition of the correlator (\ref{eq:Grho}), as
\bea 
\label{eq:rho_estimation}
\widehat\rho(\omega_0) &=& \int_0^\infty d\om\, \sum_n q_n(\omega_0) K(\tau_n,\om)\rho(\om) \nn\\
&=&  2\pi\sum_n q_n(\omega_0) G(\tau_n).
\eea
The conductivity is then extracted in a usual way,
\be
\label{eq:BG_final}
\sigma \sim  \lim_{\omega_0 \rightarrow 0} \frac{\widehat\rho(\omega_0)}{\omega_0},
\ee
assuming that the estimator $\widehat\rho(\om_0\sim 0)$ is close to the physical spectral function.
The goal is therefore to find the optimal set of functions $q_n(\om_0)$ which make $\delta(\omega_0,\omega)$ as narrow as possible in $\om$ for $\omega_0 \sim 0$.

The main difficulty in this formulation is the inversion of the matrix $W(\omega_0)$, which is usually ill-conditioned, due to the exponential decay of the kernel. This can be ameliorated by rescaling, see Eq.\ (\ref{eq:rescale}), changing the estimator to
\be
\widehat \rho(\omega_0) = f(\omega_0) \int_0^\infty d \omega\, \delta(\omega_0, \omega) \frac{\rho(\omega)}{f(\omega)},
\ee
and finding an optimal choice for $f(\om)$ (e.g.\ $f(\om) = \om$). Secondly, the matrix $W$ can be regularised as
\begin{equation}
\label{eq:W_regularised_lambda}
W_{nm} \rightarrow \lambda W_{nm} + (1 - \lambda) S_{nm},
\end{equation}
where $S_{nm}$ is the covariance matrix for the correlator $G(\tau_n)$, and $\lambda$ is a tunable parameter, determined by comparing the behaviour of $\delta(\om_0, \om)$ for different values of $\lambda$. Further discussion on successes and limitations of this approach in the context of the conductivity can be found in Refs.\ \cite{Brandt:2015aqk,Astrakhantsev:2019zkr}.

\subsection{Tikhonov regularisation}

The method of additive regularisation, see Eq.~(\ref{eq:W_regularised_lambda}), is not the only one. Tikhonov regularisation~\cite{ATikhonov:reg} acts as a complimentary instrument to other reconstruction methods, where an inversion of an ill-conditioned matrix has to be performed. Let us consider the $N\times N$ matrix $W$, defined in Eq.~(\ref{eq:W}), for which a straightforward inversion fails. A singular value decomposition yields 
\begin{eqnarray}
W &=& U \Sigma V^T,\quad U^T U = V^T V = \id, \label{eq:SVD}\\
\Sigma &=& \mbox{diag}(\sigma_1,\,\sigma_2,\,\ldots,\,\sigma_N), \nonumber
\end{eqnarray}
with $\sigma_1 \geq \sigma_2 \geq \ldots \geq \sigma_N$. Since $W^{-1} = V \Sigma^{-1} U^T$, the inversion of $W$ comes down to the inversion of $\Sigma$, which can be regularised as follows,
\begin{equation}
\label{eq:sigma_reg_ATikhonov}
\Sigma^{-1} = \mbox{diag}\left( \frac{\sigma_1}{\sigma_1^2 + \eps^2},\,\frac{\sigma_2}{\sigma_2^2 + \eps^2},\,\ldots,\,\frac{\sigma_N}{\sigma_N^2 + \eps^2} \right),
\end{equation}
where the parameter $\eps$ has to be chosen carefully; small $\eps$'s lead to precise but unstable results, while large $\eps$'s guarantee stable inversion at a cost of loss of accuracy. Further experiments with this approach can be found in Ref.\ \cite{Astrakhantsev:2019zkr}.

\subsection{Other approaches}

Here we briefly list some additional inversion methods.
Refs.\ \cite{Burnier:2011jq,Burnier:2012ts} further develop the proposal of Ref.~\cite{Cuniberti}, which formulates a unique analytic continuation which can be constructed explicitly, provided the correlator satisfies certain asymptotic behaviour in Minkowski time. The crucial requirement is that a continuum-extrapolated result for the lattice correlator is available, and that short-distance divergences, present at zero temperature, have been subtracted. So far these requirements are only met in quenched QCD; in Sec.\  \ref{sec:lattice} the application of this approach will be discussed further.
  
The following methods have not been yet been applied to the determination of the conductivity, or other transport coefficients, in QCD, as far as we know. 
The Maximum Entropy Method is only one of a number of Bayesian methods; alternative Bayesian approaches can be found in Refs.\ \cite{Burnier:2013nla,Rothkopf:2011ef}.
Ref.\ \cite{Tripolt:2018xeo} proposes an inversion based on the Schlessinger point or Resonances Via Pad\'e method, which is based on a rational-fraction representation similar to Pad\'e approximation methods. It is found that the method is competitive to MEM and Backus-Gilbert, provided the errors of the input data are small enough. Interestingly, Ref.\ \cite{Tripolt:2018xeo} applies the method to the extraction of the conductivity in graphene, described  by a tight-binding model. 

A very recent development is the implementation of machine learning approaches to tackle spectral reconstruction. Supervised learning of fully connected as well as convolutional neural networks was applied to mock data in Ref.\  \cite{LKades2019spectral}.
In Ref.\ \cite {Arsenault} kernel ridge regression was applied to mock data in quantum many-body physics; a first application to QCD data can be found in Ref.\ \cite{Offler:2019eij} for bottomonium correlators. 
More developments in the realm of machine learning are expected in the near future. 

%-----------------------------------------------------------------------------------------------------

\section{Lattice QCD results}
\label{sec:lattice}

In this section we give an overview of results obtained for the electrical conductivity in QCD, with gauge group SU(3), for $N_f=0$ (quenched QCD) and $N_f=2$ and $2+1$ dynamical flavours. We do not discuss results obtained in effective models or in other gauge theories; see Sec.~\ref{sec:related} for some results in the SU(2) theory. The papers we discuss are listed in Table \ref{tab:physical_setup_summary_table} in chronological order, with some details on the ensembles. Refs.~\cite{Gupta:2003zh,Aarts:2007wj,Ding:2010ga,Francis:2011bt,Ding:2016hua} concern quenched QCD, while in Refs.~\cite{Brandt:2012jc,Amato:2013naa,Aarts:2014nba,Brandt:2015aqk,Astrakhantsev:2019zkr} dynamical quarks are included ($N_f=2$ and $2+1$). In the dynamical studies the sea quarks are of the Wilson-clover type, with a pion mass heavier than in nature.  The exception is Ref.~\cite{Astrakhantsev:2019zkr}, with staggered sea quarks and a physical pion mass. All studies employ an isotropic lattice, with $a_s=a_\tau$ (the spatial and temporal lattice spacing respectively), except Refs.~\cite{Amato:2013naa,Aarts:2014nba}, in which anisotropic lattices, with $a_s/a_\tau=3.5$, are employed. The advantage of the latter is the finer temperature resolution, when $N_\tau$ is varied.
The lattice cutoff is dealt with in a variety of ways:
\begin{itemize}
    \item continuum limit, indicated by $a_\tau\to 0$;
    \item fixed scale: the lattice cutoff is fixed and temperature is varied by changing $N_\tau$, according to $T=1/(a_\tau N_\tau)$; \item fixed cutoff: one lattice spacing is available at each temperature, lattice spacings at different temperatures are not identical.
\end{itemize}
In simulations with dynamical quarks, continuum limits are not yet available and the (temporal) lattice spacings lie in the range $0.0350< a_\tau < 0.0618$ fm. 

%------------------ start tables -----------------------------------------------------------------------------------

\begin{table*}
\begin{center}
\caption{Details of the lattice QCD ensembles to compute the electrical conductivity. Here $a_\tau$ and $a_s$ denote the temporal and spatial lattice spacing respectively.}
\label{tab:physical_setup_summary_table}
\begin{tabular}{c c c c c c c c }
\hline\noalign{\smallskip}
ref.\ & arXiv number & $N_f$ (sea) & fermion type  & $m_\pi$ [MeV] & $a_\tau$[fm] & $a_s/a_\tau$  & discretisation   \\
\noalign{\smallskip}\hline\noalign{\smallskip}

\cite{Gupta:2003zh}	&  \href{https://arxiv.org/abs/hep-lat/0301006}{hep-lat/0301006} & 0 & quenched & $-$ & $a_\tau\to 0$ & 1 & continuum limit \\

\cite{Aarts:2007wj}	& \href{https://arxiv.org/abs/hep-lat/0703008}{hep-lat/0703008} & 0 & quenched & $-$ & $0.0488, 0.0203$ & 1 & fixed cutoff\\

\cite{Ding:2010ga}	&  \href{https://arxiv.org/abs/1012.4963}{1012.4963} & 0 & quenched & $-$ & $a_\tau\to 0$ & 1 & continuum limit\\
 
\cite{Francis:2011bt}	&  \href{https://arxiv.org/abs/1112.4802}{1112.4802} & 0 & quenched & $-$ & $0.015$ & 1 & fixed scale\\
  
\cite{Brandt:2012jc}	&  \href{https://arxiv.org/abs/1212.4200}{1212.4200} & 2 & Wilson-clover & 270 & $0.0486(4)(5)$ & 1 & fixed cutoff \\
 
\cite{Amato:2013naa}&   \href{https://arxiv.org/abs/1307.6763}{1307.6763} & $2+1$ & Wilson-clover & 384(4) &$0.0350(2)$ & 3.5 & fixed scale \\

\cite{Aarts:2014nba}	&  \href{https://arxiv.org/abs/1412.6411}{1412.6411} & $2+1$ & Wilson-clover & 384(4) & $0.0350(2)$& 3.5 & fixed scale \\
 
\cite{Brandt:2015aqk}&   \href{https://arxiv.org/abs/1512.07249}{1512.07249} & $2$ & Wilson-clover & 270 & $0.0486(4)(5)$ & 1 & fixed scale\\
 
\cite{Ding:2016hua}	&   \href{https://arxiv.org/abs/1604.06712}{1604.06712} & 0 & quenched & $-$ & $a_\tau\to 0$ & 1 & continuum limit\\

\cite{Astrakhantsev:2019zkr} &  \href{https://arxiv.org/abs/1910.08516}{1910.08516} & $2+1$ & staggered & 134.2(6) & $0.0618, 0.0493$ & 1 & fixed cutoff\\
 
\noalign{\smallskip}\hline
\end{tabular}
\end{center}
\end{table*}

\begin{table*}[t]
\begin{center}
\caption{Lattice sizes used compute the electrical conductivity: number of spatial ($N_s$) and temporal ($N_\tau$) lattice points, and corresponding temperatures. Details of the so-called ``zero-temperature'' lattices used for tuning are not listed.}
\label{tab:sizes_and_temp_summary}    
\begin{tabular}{c c c c }
\hline\noalign{\smallskip}
ref.\ & $N_s$ & $N_\tau$ & temperature \\
\noalign{\smallskip}\hline\noalign{\smallskip}

\cite{Gupta:2003zh}	& 18, $\ldots$, 44 & 14, 12, 10, 8 & $T/T_c=1.5, 2, 3$\\

\cite{Aarts:2007wj}	& 48, 64 & 24, 16 & $T/T_c=0.62, 1.5, 2.25$  \\

\cite{Ding:2010ga}	& 128 & 48, 32, 24, 16 & $T/T_c=1.45$  \\
 
\cite{Francis:2011bt}	& 128 & 40, 32, 16 & $T/T_c=1.16, 1.49, 2.98$ \\
  
\cite{Brandt:2012jc}	& 64 & 16 & $T=250$ MeV   \\
 
\cite{Amato:2013naa, Aarts:2014nba}& 24, 32 & 48, 40, 36, 32, 28, 24, 20, 16 & $T=117, 141, 156, 176, 201, 235, 281, 352$ MeV \\

\cite{Brandt:2015aqk} & 64 & 24, 20, 16, 12 & $T=169, 203, 254, 338$ MeV \\
 
\cite{Ding:2016hua}	& 96, 128, 144, 192 & 64, 56, 48, 42, 32, 28, 24 & $T/T_c = 1.1, 1.3, 1.5$ \\

\cite{Astrakhantsev:2019zkr} & 48, 64 & 10, 16 & $T=200, 250$ MeV \\
 
\noalign{\smallskip}\hline
\end{tabular}
\end{center}
\end{table*}

\begin{table*}
\begin{center}
\caption{Details of the current and inversion method used to compute the electrical conductivity.}
\label{tab:currents_and_methods_summary}
\begin{tabular}{c c c c c c c }
\hline\noalign{\smallskip}
ref.\  &   fermion type & current & renormalised & inversion method   \\
\noalign{\smallskip}\hline\noalign{\smallskip}

\cite{Gupta:2003zh}	& staggered & local & $-$ & Bayesian priors, MEM, Ansatz \\

\cite{Aarts:2007wj}	& staggered & local & $-$ & MEM  \\

\cite{Ding:2010ga}	& Wilson-clover & local & \checkmark & Ansatz, MEM \\
 
\cite{Francis:2011bt} & Wilson-clover & local & \checkmark & Ansatz, MEM \\
  
\cite{Brandt:2012jc} & Wilson-clover & local & \checkmark & Ansatz, sum rule constraints \\
 
\cite{Amato:2013naa,Aarts:2014nba}& Wilson-clover & conserved & \checkmark & MEM \\

\cite{Brandt:2015aqk}& Wilson-clover & mixed local-conserved & \checkmark & Ansatz, sum rule constraints, Backus-Gilbert \\
 
\cite{Ding:2016hua} & Wilson-clover & local & \checkmark & Ansatz \\

\cite{Astrakhantsev:2019zkr} & staggered & conserved & \checkmark & Backus-Gilbert, Tikhonov regularisation\\
 
\noalign{\smallskip}\hline
\end{tabular}
\end{center}
\end{table*}

%------------------ end tables -----------------------------------------------------------------------------------

Table~\ref{tab:sizes_and_temp_summary} contains the details of the lattice geometry, i.e.\ the number of lattice points in spatial ($N_s$) and temporal ($N_\tau$) direction. The corresponding temperatures are expressed in units of $T_c$ for the quenched ensembles and in units of MeV for the dynamical ones. The largest number of temperatures is considered in Refs.~\cite{Amato:2013naa,Aarts:2014nba}, with 8 temperature values, ranging from 117 to 352 MeV. 

Table~\ref{tab:currents_and_methods_summary} finally lists some details on the currents and methods used to compute the conductivity:
\begin{itemize}
    \item so far the type of valence quarks used in the current have been either staggered or Wilson-clover fermions. For staggered quarks, the current-current correlator (\ref{eq:Gem}) has an oscillating structure, of the form \cite{Aarts:2007wj,Astrakhantsev:2019zkr},
    \begin{eqnarray}
    G_{ij}(x) &=& \langle A_i(x) A_j(0) \rangle - (-1)^{\tau/a_\tau} \langle B_i(x) B_j(0) \rangle, \nn \\
    A_i &=& \bar \psi\gamma_i \psi, \qquad\quad  B_i = \bar \psi \gamma_4 \gamma_5 \gamma_i  \psi.
    \end{eqnarray}
    To resolve this staggering, the spectral reconstruction is performed on even and on odd time slices independently, obtaining two spectral functions, $\rho^{\rm even, odd}_{ij}(\omega)$. In the sum
    \begin{equation}
    \rho_{ij}(\omega) = \half\left[ \rho^{\rm even}_{ij}(\omega) + \rho^{\rm odd}_{ij}(\omega) \right],
    \end{equation}
    the oscillating contribution cancels, and one can proceed with $\rho_{ij}(\omega)$ as usual. This procedure limits, however, the number of time slices effectively available by a factor of two. This issue does not arise with Wilson-type fermions, which are hence the preferred choice.

    \item local/conserved current: the simplest choice for the current operator is the local one, $j^{\rm loc}_{x,i} = \bar\psi_x \gamma_i\psi_x$, with the quark fields residing at the same lattice point $x$. This operator requires renormalisation, i.e.\ the determination of a renormalisation factor $Z_V$. The earliest contributions \cite{Gupta:2003zh,Aarts:2007wj} did not determine this renormalisation factor, which was accounted for in the systematic uncertainty. Later studies using the local current did renormalise it properly. 
    
    The conserved current, of the point-split form (the expression below is for Wilson fermions, $U_{x,i}$ is the gauge link in the spatial direction)
    \be
    j^{\rm cons}_{x,i} = \half\bar\psi_{x+\hat\imath}(1+\gamma_i) U^\dagger_{x,i}\psi_x - \half\bar\psi_x(1-\gamma_i) U_{x,i}\psi_{x+\hat\imath},
    \ee
    is the Noether current corresponding to a global phase symmetry of the fermion lattice action and hence does not require renormalisation, even at finite lattice spacing. For the current-current correlator, $\bra j_{x,i}j_{y,i}\ket$,  combining two conserved currents, as in Refs.\ \cite{Amato:2013naa,Aarts:2014nba,Astrakhantsev:2019zkr},
    eliminates the need to compute the $Z_V$ factor, but it is also the most expensive numerically, due to the need to invert more combinations of quark propagators. Ref.~\cite{Brandt:2015aqk} employs a mixed combination, $\bra j^{\rm cons}_{x,i}j^{\rm loc}_{y,i}\ket$, which is cheaper to evaluate and exactly conserved on one side. It still requires knowledge of the $Z_V$ factor, which can e.g.\ be obtained from 
    $\bra j^{\rm loc}_{x,i}j^{\rm loc}_{y,i}\ket/\bra j^{\rm cons}_{x,k}j^{\rm loc}_{y,k}\ket$.

    \item the final column lists the methods, discussed above, employed to reconstruct the spectral function and extract the conductivity. Ans\"atze and MEM have traditionally been the most popular ones, with the Backus-Gilbert method and Tikhonov regularisation being applied to this problem more recently.

\end{itemize}

Before moving to the results obtained so far, we note that none of the papers listed above include the so-called disconnected contributions, discussed in Sec. \ref{sec:kubo}. 
The neglect of the disconnected contribution can be motivated in a number of ways:
\begin{itemize}
    \item the contribution vanishes at very high $T$, since it is ${\cal O}(\alpha_s^3)$ at leading order in perturbation theory;
    \item the contribution vanishes for three degenerate flavours, due to the sum over the charges;
    \item the contribution is expected to be noisy numerically. 
\end{itemize}
This last comment is an excuse, rather than a motivation, and indeed, it would be of interest to estimate the level of statistics required to compute a signal in the $N_f=2$ or the non-degenerate $N_f=2+1$ case and verify e.g.\ the first remark at high temperature.
We also note that all studies discussed here have been carried out at zero spatial momentum; the extension to nonzero momentum is straightforward (see e.g.~Ref.~\cite{Aarts:2006cq}) and can provide an additional handle on hydrodynamic behaviour at small $\om$ and $|\vecp|$. 

After this overview of the lattice details, we are now in a position to compare the conductivities computed in the references listed above. The conductivity is normalised with the temperature (to make it dimensionless) and with $C_{\rm em} = \sum_f (eq_f)^2$, the sum over the charges squared, see Eq.\ (\ref{eq:Cem}). The latter division allows one to compare e.g.\ the $N_f=2$ and $2+1$ cases. We separately discuss the quenched results and the results with dynamical quarks.

\begin{figure}[t]
    \centering
    \includegraphics[scale=0.55]{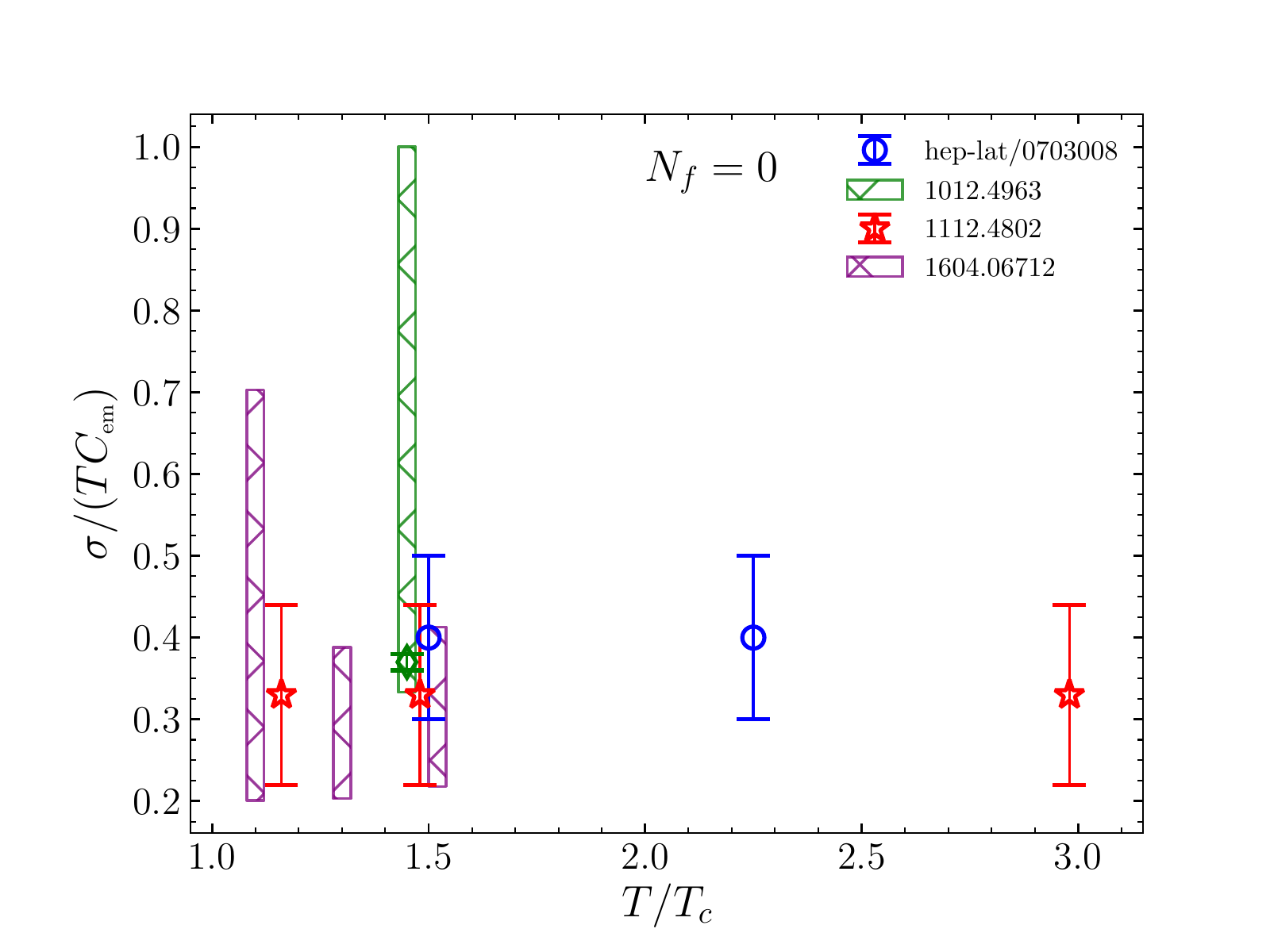}
    \caption{Results for the electrical conductivity, normalised as $\sigma/(TC_{\rm em})$, in quenched QCD ($N_f=0$) as a function of $T/T_c$. 
    The early result \cite{Gupta:2003zh}, $\sigma/(T C_{\rm em}) \approx 7$, is not shown for clarity.}
    \label{fig:sigma_gluo_summary}
\end{figure}

\begin{table}[b]
\begin{center}
\caption{Estimates of the pseudocritical temperatures for the studies with dynamical quarks.}
\label{tab:Tc}
\begin{tabular}{c c c c }
\hline\noalign{\smallskip}
ref.\  &   $N_f$ & $m_\pi$ [MeV] & $T_{\rm pc}$ [MeV]   \\
\noalign{\smallskip}\hline\noalign{\smallskip}

\cite{Brandt:2012jc,Brandt:2015aqk} & $2$ & $270$ & $\sim 203$  \\
\cite{Amato:2013naa,Aarts:2014nba}& $2+1$ & $384(4)$ & $185(4)$ \\
\cite{Astrakhantsev:2019zkr} & $2+1$ & $134.2(6)$ & $155(2)(3)$ \\
 
\noalign{\smallskip}\hline
\end{tabular}
\end{center}
\end{table}

Results in quenched QCD are shown in Fig.~\ref{fig:sigma_gluo_summary}, as a function of $T/T_c$. The very early result from Ref.~\cite{Gupta:2003zh}, $\sigma/(T C_{\rm em}) \approx 7$, is not included, since it is about a factor of $20$ larger than the other results. The remaining four quenched studies are in good agreement, with a value of $\sigma/(T C_{\rm em}) \approx 0.2-0.5$. Note that Ref.\ \cite{Ding:2010ga} provides both a precise result, $\sigma/(TC_{\rm em})= 0.37(1)$ at $T/T_c=1.45$, and a more conservative range, indicated with the tallest vertical green column.
Although these are four studies, we note that they emerge from two groups only, Ref.~\cite{Aarts:2007wj} on the one hand and Refs.~\cite{Ding:2010ga,Francis:2011bt,Ding:2016hua} on the other hand, making the agreement is perhaps less surprising. In any case, it is interesting that the early quenched result of Ref.~\cite{Aarts:2007wj}, obtained using staggered quarks without taking a continuum limit, remains to be consistent with the renormalised continuum-extrapolated Wilson-clover results of  Refs.~\cite{Ding:2010ga,Francis:2011bt,Ding:2016hua}.
A second observation of interest is that there appears to be very little temperature dependence in the temperature range investigated, $1.1< T/T_c<3$.
We remind the reader that in quenched QCD the deconfinement transition is first-order, signalled by the spontaneous breaking of the centre symmetry. 

\begin{figure}[t]
    \centering
    \includegraphics[scale=0.55]{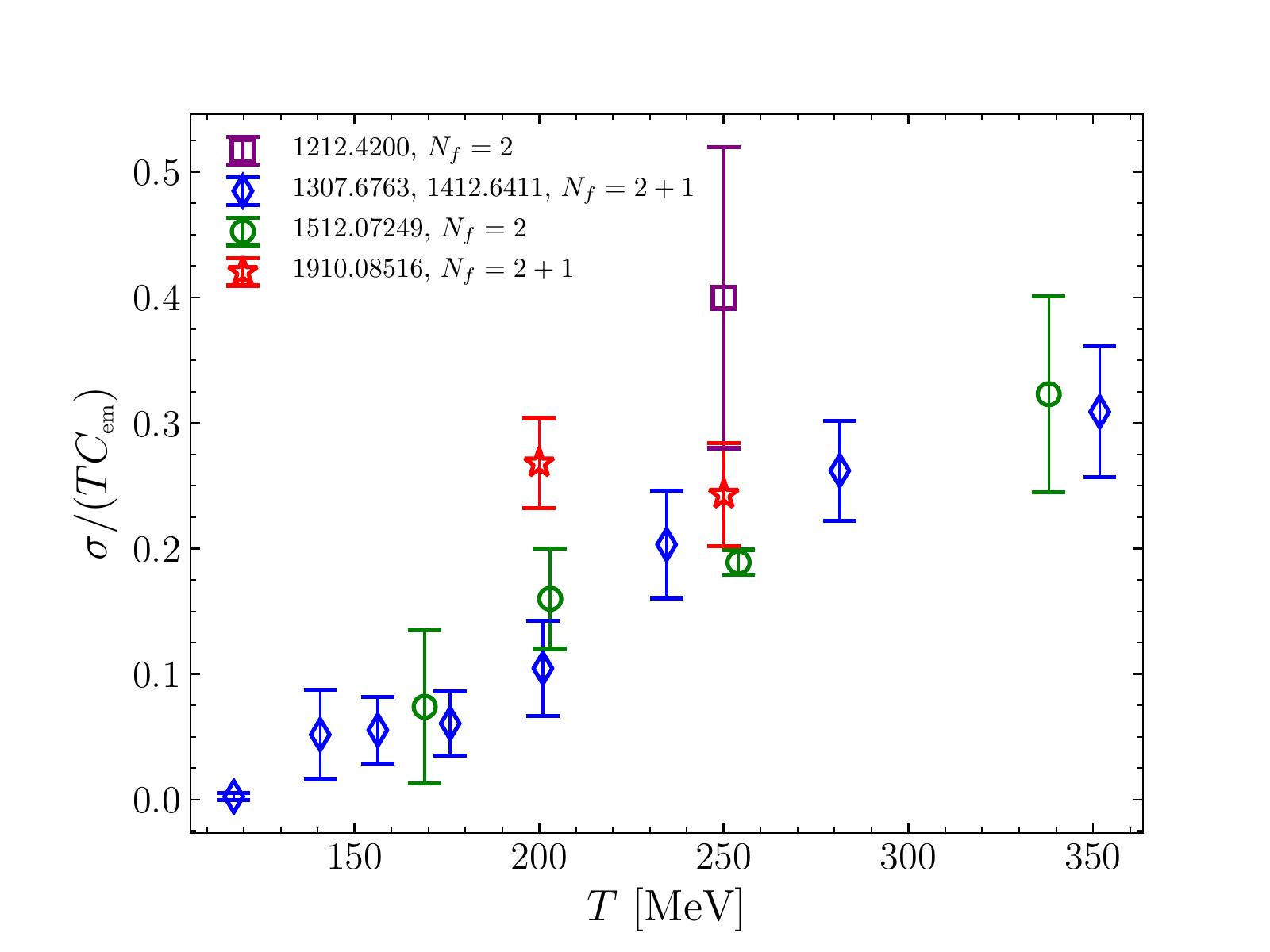}
    \caption{Results for electrical conductivity, normalised as $\sigma/(TC_{\rm em})$,  in QCD with $N_f=2$ and $2+1$ dynamical flavours as a function of temperature in MeV. 
    }
    \label{fig:sigma_ferm_summary}
\end{figure}

We now turn to the dynamical results, shown in Fig.~\ref{fig:sigma_ferm_summary} as a function of temperature in MeV. In this case, the thermal transition is a smooth crossover. It should be noted that the crossover temperature is slightly different between the various studies, due to the difference in the pion mass and the number of flavours, see Table \ref{tab:Tc} (note that Ref.\ \cite{Astrakhantsev:2019zkr} follows the lattice formulation and choice of parameters of Refs.\ \cite{Aoki:2009sc,Borsanyi:2010cj}. Moreover, Ref.~\cite{Astrakhantsev:2019zkr} has results at two lattice spacings; the red star symbols denote the results at the smaller spacing, $N_\tau = 16$).
In particular, the transition temperatures in Refs.~\cite{Amato:2013naa, Aarts:2014nba,Brandt:2015aqk} are higher than in Ref.\ \cite{Astrakhantsev:2019zkr}, with simulations in the latter being at the physical point. 
In Fig.~\ref{fig:sigma_ferm_summary} a temperature-dependent $\sigma/T$ can be observed, with a reduction in the vicinity of the thermal crossover, in contrast to the quenched case. This temperature dependence is especially visible in the data from Refs.~\cite{Amato:2013naa, Aarts:2014nba} and Ref.~\cite{Brandt:2015aqk}, which have results at a number of temperatures (8 and 4 values respectively). The difference in crossover temperatures might explain the slightly lower values for the conductivity at $T=200$ MeV in Refs.~\cite{Amato:2013naa, Aarts:2014nba,Brandt:2015aqk}, compared to Ref.~\cite{Astrakhantsev:2019zkr}, as in the former cases 200 MeV is closer to the pseudocritical temperatures and the thermal crossover region. As mentioned above, in quenched QCD, with a first-order transition, no reduction of the conductivity above the critical temperature is seen. These observations suggest that the reduction of the conductivity is due to the smooth transition to the hadronic phase. 
We also note that Refs.\ \cite{Brandt:2012jc,Brandt:2015aqk} are from the same group; hence the data point indicated by the square is effectively superseded by those indicated with the circles. 
This allows us to conclude that there is good consistency between the various studies, which is a nontrivial result, given the difference in lattice formulations, lattice geometries and inversion methods employed. Finally we observe that at the highest temperatures studied the value for $\sigma/(TC_{\rm em})$ is comparable to the one obtained in the quenched case.

In order to judge whether the observed magnitude of the conductivity signifies strong- or weak-coupling behaviour, we note that at weak coupling (including only QCD processes) the usual expectation is that $\sigma/(TC_{\rm em})\sim 1/g^4\ln 1/g$ \cite{Arnold:2000dr}, which is much larger than 1 in the region where the weak-coupling analysis is valid, namely at asymptotically high temperatures. A useful benchmark at strong coupling comes from holography, for the charge diffusion coefficient $D=\sigma/\chi_Q$, where $\chi_Q$ is the charge susceptibility. The characteristic result at strong coupling is $D=1/(2\pi T)$ in ${\cal N} = 4$ Yang- Mills theory at nonzero temperature \cite{Policastro:2002se,Teaney:2006nc}. In Ref.~\cite{Aarts:2014nba} the temperature dependence of $D$ was computed in a self-contained manner, i.e.\ by also computing $\chi_Q$ within the same lattice QCD setup, with the result that $0.5<2\pi TD<2$, compatible with the holographic order of magnitude at strong coupling. Moreover, it was observed that $2\pi TD$ has a minimum in the crossover region, see Fig.~14 of Ref.~\cite{Aarts:2014nba}. 

Before concluding this section, we note that the lattice data of Ref.~\cite{Ding:2010ga} have been re-analysed in two papers. Ref.~\cite{Burnier:2012ts} employed the approach of Ref.~\cite{Cuniberti}, see also Ref.~\cite{Burnier:2011jq}, in which short-distance divergences are subtracted from the Euclidean correlator. Additional insight on the ultraviolet asymptotics of the thermal contribution to the spectral function is taken from Ref.~\cite{CaronHuot:2009ns}, such that only the contribution of the vacuum spectral function needs to be subtracted. For this, a 5-loop computation of the vector current correlator in vacuum is employed. A smaller result for the conductivity and diffusion coefficient are found with respect to Ref.~\cite{Ding:2010ga}, namely, at $T/T_c=1.45$,
\be
\label{eq:ML}
 \sigma/(T C_{\rm em}) \gtrsim ~ 0.1, \qquad\quad 2\pi T D \gtrsim 0.8,
 \ee
which is indeed smaller by a factor of 3 for the conductivity. It is stated \cite{Burnier:2012ts} that the results in Eq.~(\ref{eq:ML}) should be interpreted as lower bounds. 

The data of Refs.~\cite{Ding:2010ga} and \cite{Brandt:2015aqk} has also been re-analysed in Refs.~\cite{Gubler:2016hnf} and \cite{Gubler:2017qbs} respectively, using thermal sum rules to constrain the Ans\"atze used in the fits. In Ref.~\cite{Gubler:2016hnf}, a higher result was found for quenched study at $T/T_c=1.45$, namely 
\be
 \sigma/(T C_{\rm em}) \sim 0.57.
 \ee
In Ref.~\cite{Gubler:2017qbs} the $N_f=2$ data given in Ref.~\cite{Brandt:2015aqk} at four temperatures was re-analysed.
Approximate agreement was found, with the important caveat that the fits were seen not to be as stable as desired.

Even though good consistency between the various studies can be seen, nevertheless continuing uncertainty in spectral reconstruction remains, with an ongoing need to further develop methods for analytical continuation, emphasising robustness and quantification of underlying uncertainties.

%-----------------------------------------------------------------------------------------------------

\section{External conditions}
\label{sec:related}

The electrical conductivity, as well as other transport coefficients, may be studied under other external conditions than temperature, such as in the presence of an external magnetic field or at nonzero quark density.
In this section we briefly mention some related developments in nonabelian gauge theories.

The first attempts to investigate the dependence of the electrical conductivity on an external magnetic field using lattice simulations were performed in the quenched SU(2) theory in Refs.~\cite{Buividovich:2010tn,Buividovich:2010qe}, using the overlap Dirac operator with exact chiral symmetry  in the current-current correlator. MEM is used for spectral reconstruction. The emphasis is on the conductivity in the presence of an external magnetic field, both in the confined and the deconfined phase, and on the quark mass dependence. It is found that in the confined phase the external magnetic field induces a nonzero electric conductivity along the direction of the field, while in the deconfined phase no sizable dependence on the magnetic field is observed.

In Ref.~\cite{Buividovich:2020el} the conductivity is studied in the SU(2) gauge theory with dynamical quarks at nonzero density, which is feasible due the absence of a sign problem in this theory. Gauge configurations are generated with dynamical staggered quarks, while current-current correlators are computed with Wilson-Dirac and Domain Wall fermions, tuned in such a way to match the pion mass of the ensembles. The conductivity is extracted via several methods, including the Backus-Gilbert method with the use of Tikhonov regularisation. At small quark chemical potential $\mu$, the dependence on chemical potential is considered via the expansion
\be
\frac{\sigma(\mu)}{\sigma(0)} =  1+ c(T)\frac{\mu^2}{T^2} + {\cal O}\left( \frac{\mu^4}{T^4}\right),
\ee
and it is found that the maximal value of the second-order coefficient, $c(T)\approx 0.15(5)$, is reached in the vicinity of the chiral crossover. Hence the coefficient $c(T)$ is quite small, and even at $\mu/T \approx 1$ the conductivity changes no more than 15--20\% compared to its zero-density value. As for the large density region, QCD with $N_c = 2$ and $3$ colours differ, due to the formation of a diquark condensate in the former. Nevertheless, at smaller $\mu$ the SU(2) theory can provide qualitative insights for real QCD.

Also in QCD (with $N_c=3$ and $N_f = 2 + 1$) the electrical conductivity has been studied in the presence of an external constant uniform magnetic field~\cite{Astrakhantsev:2019zkr}. The $\vecB=\vecnul$ results of this reference have been discussed above; with nonzero $\vecB$ field, it is found that the conductivity rises in the direction parallel to the magnetic field and decreases in the transverse direction. This may potentially be explained by the Chiral Magnetic Effect~ \cite{Kharzeev:2015znc} and magnetoresistance.

%-----------------------------------------------------------------------------------------------------

\section{Conclusion}
\label{sec:concl}

In this paper we reviewed the status of the electrical conductivity in the quark-gluon plasma, as seen through nonperturbative lattice QCD simulations. After an overview of basic definitions and expectations, we listed several methods that have been used for spectral reconstruction, the main challenge in this endeavour. No method has yet reached full acceptance, due to the apparent lack of robustness and handle on systematic uncertainties. This remains therefore the outstanding challenge to be tackled. It is also noted that none of the results include the disconnected contributions yet, for reasons discussed in Sec.~\ref{sec:lattice}. It would be worthwhile to estimate the importance of those eventually.

Nevertheless, a comparison between the existing lattice studies, presented in Sec.~\ref{sec:lattice}, reveals a noticeable consistency, which is encouraging, given the difference in lattice formulations, lattice geometries (in particular the number of temporal points), and %% change suggestion
reconstruction methods employed. Taking the results at face value, the main findings are 
\begin{itemize}
    \item in quenched ($N_f=0$) QCD $\sigma/T$ appears to have very little temperature dependence in the temperature range investigated, $1.1< T/T_c<3$. The magnitude is approximately $0.2 \lesssim \sigma/(T C_{\rm em}) \lesssim 0.5$, where $C_{\rm em}$ is the sum over electric charges squared appearing in the electromagnetic current-current correlator;
    \item in QCD with $N_f=2$ and $2+1$ dynamical flavours, the main finding is a noticeable reduction of $\sigma/T$ in the vicinity of the thermal crossover, compared to its value at higher temperatures in the QGP. This should be contrasted with the quenched case. 
    This effect has been observed by two groups independently and is further (indirectly) supported by simulations at the physical point by a third group. One possible interpretation is that the reduction of the conductivity is due to the smooth transition to the hadronic  phase. 
    This might be of interest for phenomenology.
    It is further noted that at the highest temperatures studied the value for $\sigma/(T C_{\rm em})$ is comparable to the one obtained in the quenched case. Overall, the magnitude of the conductivity is compatible with the plasma being strongly coupled, using the comparison of the charge diffusion coefficient $D=\sigma/\chi_Q$, where $\chi_Q$ is the charge susceptibility, with the one obtained in holography.
\end{itemize}

So far most studies have focused on the quark-gluon plasma and the crossover region. Deeper in the hadronic phase the conductivity should be dominated by the lightest charged hadrons. So far, lattice studies have not given a detailed study of this regime, possibly because the signal is hard to detect. We refer to Ref.~\cite{FernandezFraile:2009mi} for an overview from the perspective of chiral perturbation theory. Studies in the presence of an external magnetic field or at finite density require further attention. While direct access to the latter is not feasible in QCD due to the sign problem, a Taylor series expansion in the powers of $\mu/T$ is possible, although it is expected to be noisy numerically~\cite{Aarts:2020lvh}. Another interesting possibility is the analysis of the current-current correlator at imaginary chemical potential.

%-----------------------------------------------------------------------------------------------------

\begin{acknowledgement}

We thank Mikko Laine and Pavel Buividovich for discussion. GA is grateful for collaboration with Chris Allton, Ale Amato, Justin Foley, Pietro Giudice, Simon Hands, Seyong Kim, and Jon-Ivar Skullerud on some of the results reviewed here. This work is supported by STFC via grant ST/P00055X/1, the European Research Council (ERC) under the European Union's Horizon 2020 research and innovation programme via grant agreement No.~813942, and by COST Action CA15213 THOR. A.N.~also acknowledges the support by RFBR grant No.~18-02-40126.

\end{acknowledgement}

\appendix

\section{Green's functions}
\label{app:A}

For the convenience of the reader we collect in this Appendix some relations between the various two-point functions for an operator $O(x) = O(t_x,\xv)$. These relations are well known \cite{LeBellac,Laine:2016hma}.

Let us start with the retarded and advanced Green's functions
\be
G_R(x-y) =  i\theta(t_x-t_y)\bra [ O(x), O^\dagger(y)]\ket
= G_A(y-x),
\ee
and the spectral function
\bea
\rho(x-y) &=& \bra [O(x), O^\dagger(y)]\ket
\nn\\
&=& -i \left[ G_R(x-y) - G_A(x-y) \right]. 
\eea
Expectation values are taken in thermal equilibrium, which explains the $x-y$ dependence.
After going to  momentum space and using the identity
\be
\int_{-\infty}^\infty \frac{d\om}{2\pi} \, \frac{e^{-i\om t}}{\om+i\epsilon} = -i\theta(t),
\ee
we arrive at the dispersion relation
\be
G_R(\om,\pv) = \int_{-\infty}^\infty \frac{d\om'}{2\pi} \, \frac{\rho(\om',\pv)}{\om'-\om-i\epsilon}.
\ee
Employing the identity
\be
\frac{1}{x+i\epsilon}-\frac{1}{x-i\epsilon} = \frac{-2i\epsilon}{x^2+\epsilon^2}\to -2i\pi\delta(x),
\ee
then yields the important relation,
\be
\rho(\om,\pv) = -i\left[ G_R(\om,\pv)-G_A(\om,\pv)\right] = 2\im\, G_R(\om,\pv),
\ee
i.e.\ the spectral function is twice the  imaginary part of retarded Green function, or equivalently the discontinuity across the real axis.

The Euclidean correlator, 
\be
G_E(\tau,\xv) = \bra O(\tau,\xv)O^\dagger(0,\vecnul)\ket,
\ee
with $0\leq \tau < 1/T$, is written in momentum space as
\bea
G_E(\om_n,\pv) &=& \int_0^{1/T}d\tau\, e^{i\om_n \tau}G_E(\tau,\pv), \\
G_E(\tau,\pv) &=& T\sum_n e^{-i\om_n \tau}G_E(\om_n,\pv), 
\eea
where $\om_n=2\pi nT$, $n\in \mathbb{Z}$, are the Matsubara frequencies (we consider bosonic operators here). By analyticity, it satisfies a similar dispersion relation as above,
\be
\label{eq:disp}
G_E(\om_n,\pv) = \int_{-\infty}^\infty \frac{d\om'}{2\pi} \, \frac{\rho(\om',\pv)}{\om'-i\om_n},
\ee
leading to the important relation
\be
G_R(\om,\pv) = G_E(i\om_n\to w+i\epsilon,\pv).
\ee
If a Euclidean correlator is known analytically, the spectral function can be obtained following the sequence
\be
G_E(\tau,\xv) \to G_E(\om_n,\pv) \to G_R(\om,\pv) \to \rho(\om,\pv).
\ee
Unfortunately, this path is not accessible with numerically determined correlators on a finite number of points in the temporal direction. 

Instead we will relate the correlator and the spectral function via a Laplace transform, generalised to nonzero temperature. Going back to Euclidean time, we find, using Eq.\ (\ref{eq:disp}), 
\be\label{eq:G}
G(\tau,\pv) = \int_{-\infty}^\infty \frac{d\om}{2\pi}\, \tilde K(\tau,\om)\rho(\om, \pv),
\ee
with the kernel
\be\label{eq:kernel_cosh_sinh}
\tilde K(\tau,\om) = T\sum_n \frac{e^{-i\om_n \tau}}{\om-i\om_n} = e^{-\om\tau}[1+n_B(\om)].\ee
Here $n_B(\om) = 1/[\exp(\om/T)-1]$ is the Bose-Einstein distribution and we have taken $0<\tau<1/T$.

For hermitian operators $O^\dagger(x)=O(x)$, the spectral function is odd in $\om$, 
\be
\rho(-\om,\pv) = -\rho(\om,\pv).
\ee
Hence in Eq.\ (\ref{eq:G}) only the odd part of the kernel $\tilde K(\tau, \om)$ survives, and we arrive at the standard integral relation
\be
\label{eq:AG}
G(\tau,\pv) =  \int_0^\infty \frac{d\om}{2\pi}\, K(\tau,\om)\rho(\om, \pv),
\ee
with
\bea
K(\tau,\om) &=& \tilde K(\tau,\om) - \tilde K(\tau,-\om) \nn
\\
&=&  e^{-\om\tau}[1+n_B(\om)] + e^{\om\tau}n_B(\om) \\
&=& \frac{\cosh[\om(\tau-1/2T)]}{\sinh(\om/2T)},
\eea
where we used the identity
\be
n_B(\om)+n_B(-\om)+1=0.
\ee

% BibTeX users please use
% \bibliographystyle{}
% \bibliography{}
%
% Non-BibTeX users please use

\end{document}